\documentclass[twocolumn]{revtex4}

\usepackage{amsfonts,amssymb,amsbsy,bm,graphicx}

\begin{document}

\title{Hyperbolic reflections as fundamental
building blocks for multilayer optics}

\author{Alberto G. Barriuso, Juan J. Monz\'on,
and Luis L. S\'anchez-Soto}

\affiliation{Departamento de \'Optica,
Facultad de F\'{\i}sica, Universidad Complutense,
28040 Madrid, Spain}

\author{Jos\'e F. Cari\~{n}ena}
\affiliation{Departamento de F\'{\i}sica Te\'orica,
Facultad de Ciencias, Universidad de Zaragoza,
50009 Zaragoza, Spain}

\begin{abstract}
We reelaborate on the basic properties of
lossless multilayers by using bilinear
transformations. We study some interesting
properties of the multilayer transfer
function in the unit disk, showing that
hyperbolic geometry turns out to be an
essential tool for understanding
multilayer action. We use a simple trace
criterion to classify multilayers into
three classes that represent rotations,
translations, or parallel displacements.
Moreover, we show that these three actions
can be decomposed as a product of two reflections
in hyperbolic lines. Therefore, we conclude
that hyperbolic reflections can be considered
as the basic pieces for a deeper understanding
of multilayer optics.
\end{abstract}


\maketitle

\newpage

\section{Introduction}

Although special relativity is perhaps the
first theory that comes to mind when
speaking about the interplay between physics
and geometry, one cannot ignore that geometrical
ideas are essential tools in the development of
many branches of modern physics~\cite{SC97}.

The optics of layered media is not an exception:
in recent years many concepts of geometrical
nature have been introduced to gain further
insights into the behavior of multilayers.
The algebraic basis for these developments
is the fact that the transfer matrix
associated with a lossless multilayer is
an element of the group SU(1,~1), which is
locally isomorphic to the $(2+1)$-dimensional
Lorentz group SO(2,~1).This leads to a natural
and complete identification between reflection
and transmission coefficients and the
parameters of the corresponding Lorentz
transformation~\cite{MO99a,MO99b}.

As soon as one realizes that SU(1,~1) is also
the basic group of the hyperbolic geometry~\cite{CO68},
it is tempting to look for an enriching geometrical
interpretation of the multilayer optics. Accordingly,
we have recently proposed~\cite{YO02,MO02} to view
the action of any lossless multilayer as a bilinear
(or M\"{o}bius) transformation on the unit disk,
obtained by stereographic projection of the unit
hyperboloid of SO(2,~1). This kind of representation
has been previously discussed for, e.g., the
Poincar\'e sphere in polarization optics~\cite{AZ87,HA96},
for Gaussian beam propagation~\cite{KO65}, in laser
mode-locking and optical pulse transmission~\cite{NA98},
and also in modelling visual processing~\cite{ME91}.

The point we wish to emphasize is that
these bilinear transformations preserve
hyperbolic distance between points on
the unit disk. In Euclidean geometry any
transformation of the plane that preserves
distance can be written as a composition of
reflections, which can be then considered as
the most basic transformations. In fact, the
composition of two reflections in straight lines
is a rotation, or a translation, according these
lines are intersecting or parallel.

In hyperbolic geometry, each circle orthogonal
to the boundary of the unit disk is a hyperbolic
line and reflections appear as inversions.
However, we have an essential difference with
the Euclidean case because there are three
different kind of lines: intersecting, parallel,
and ultraparallel (which are neither intersecting
nor parallel)~\cite{CO68}. In consequence, the
composition of two reflections in hyperbolic
lines is now a rotation, a parallel displacement,
or a translation: these are precisely the
transformations of the unit disk that preserve
distance.

A powerful way of characterizing transformations
is through the study of the points that they
leave invariant. For example, in Euclidean
geometry a rotation can be characterized by
having only one fixed point, while a translation
has no invariant point. For a reflection the
fixed points consist of all the points of
a line (the reflection axis).

In this paper we shall consider the fixed
points of the bilinear transformation
induced by the multilayer, showing that they
can be classified according to the trace of
the multilayer matrix. From this viewpoint,
the three transformations mentioned above;
namely, rotations, parallel displacements,
and translations appear linked to the fact
that the trace of the multilayer transfer
matrix has a magnitude lesser, equal or
greater than 2.

Since reflections appear as the basic
building blocks of these geometric
motions~\cite{DU81}, we show that any
multilayer action can be decomposed in
terms of two inversions whose meaning is
investigated. Such a decomposition is
worked out for practical examples. This
shows the power of the method and, at the
same time, allows for a deeper understanding
of layered media.

\section{Multilayers and the unit disk}

We first briefly summarize the essential
ingredients of multilayer optics we shall
need for our purposes~\cite{AZ87}. We deal
with a stratified structure, illustrated
in Fig.~1, that consists of a stack of
$1, \ldots, j, \ldots, m$, plane-parallel
layers sandwiched between two semi-infinite
ambient ($a$) and substrate ($s$) media,
which we shall assume to be identical,
since this is the common experimental
case. Hereafter all the media are supposed
to be lossless, linear, homogeneous, and
isotropic.

\begin{figure}
\centering
\resizebox{0.60\columnwidth}{!}{\includegraphics{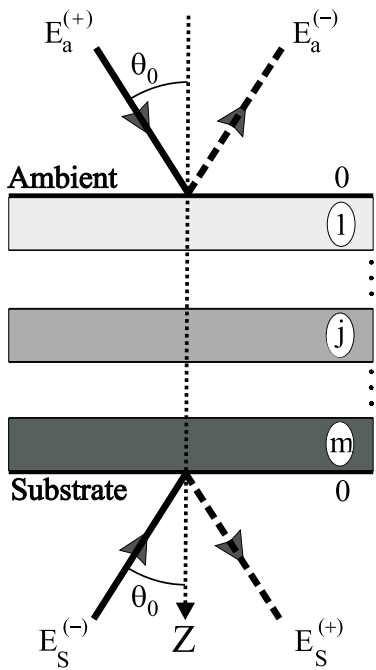}}
\caption{Wave vectors of the input $[E_{a}^{(+)}$ and
$E_{s}^{(-)}]$ and output $[E_{a}^{(-)}$ and $E_{s}^{(+)}]$ fields
in a multilayer sandwiched between two identical semi-infinite
ambient  and substrate media.}
\end{figure}

We consider an incident monochromatic linearly polarized plane
wave from the ambient, which makes an angle $\theta_0$ with the
normal to the first interface and has amplitude $E_{a}^{(+)}$. The
electric field is either in the plane of incidence ($p$
polarization) or perpendicular to the plane of incidence ($s$
polarization). We consider as well another plane wave of the same
frequency and polarization, and with amplitude $E_{s}^{(-)}$,
incident from the substrate at the same angle $\theta
_{0}$~\cite{Snell}.

As a result of multiple reflections in all
the interfaces, we have a backward-traveling
plane wave in the ambient, denoted $E_{a}^{(-)}$,
and a forward-traveling plane wave in the
substrate, denoted $E_{s}^{(+)}$. If we
consider the field amplitudes as a vector
of the form
\begin{equation}
\label{Evec}
\mathbf{E} =
\left ( \begin{array}{c}
E^{(+)} \\
E^{(-)}
\end{array}
\right )\ ,
\end{equation}
which applies to both ambient and substrate
media, then the amplitudes at each side
of the multilayer are related by a $2 \times 2$
complex matrix $\mathsf{M}_{as}$, we shall
call the multilayer transfer matrix,
in the form
\begin{equation}
\label{M1}
\mathbf{E}_a =
\mathsf{M}_{as} \
\mathbf{E}_s\ .
\end{equation}
The matrix $\mathsf{M}_{as}$ can be shown
to be~\cite{MO99a}
\begin{equation}
\label{Mlossless}
\mathsf{M}_{as} =
\left [
\begin{array}{cc}
1/T_{as} & R _{as}^\ast/T_{as}^\ast \\
R_{as}/T_{as} & 1/T_{as}^\ast
\end{array}
\right ]
\equiv
\left [
\begin{array}{cc}
\alpha & \beta \\
\beta^\ast & \alpha^\ast
\end{array}
\right ] ,
\end{equation}
where the complex numbers $R_{as}$ and
$T_{as}$ are, respectively, the overall
reflection and transmission coefficients
for a wave incident from the ambient.
Because $ |R _{as}|^2 + |T _{as}|^2 =1$,
we have the additional condition
$| \alpha |^2 - | \beta |^2 = 1$ or,
equivalently, $\det \mathsf{M}_{as}= +1$
and then the set of lossless multilayer
matrices reduces to the group SU(1,~1),
whose elements depend on three independent
real parameters.

The identity matrix corresponds to $T_{as} =1$
and $R_{as} = 0$, so it represents an
antireflection system. The matrix that
describes the overall system obtained by
putting two multilayers together is the
product of the matrices representing each
one of them, taken in the appropriate order.
So, two multilayers, which are inverse,
when composed give an antireflection
system~\cite{MO99c}.

In Refs.~[2] and [3] we have proposed to view
the multilayer action in a relativisticlike
framework. Without going into details, it is
convenient to characterize the state of the fields
at each side of the multilayer by means of the
``space-time" coordinates
\begin{eqnarray}
\label{equivalencia}
e^0 & = & \frac{1}{2}
[ |E^{(+)}|^2 + |E^{(-)}|^2] ,
\nonumber \\
e^1 & = & {\rm Re}
[{E^{(+)}}^\ast E^{(-)} ] ,
\nonumber \\
e^2 & = & {\rm Im}
[{E^{(+)}}^\ast E^{(-)} ],
\nonumber \\
e^3 & = & \frac{1}{2}
[ |E^{(+)}|^2 - |E^{(-)}|^2] ,
\end{eqnarray}
for both ambient and substrate media.
The coordinate $e^3$ is the semi-difference
of the fluxes (note that this number can
take any real value) and, therefore, is
constant because the multilayer is lossless.
In consequence, we have that
\begin{equation}
\label{intervalo}
(e^0)^2 - (e^1)^2 -
(e^2)^2 = (e^3)^2 = \mathrm{constant} .
\end{equation}
Equation (\ref{intervalo}) defines a two-sheeted
hyperboloid of radius $e^3$, which without
loss of generality will be taken henceforth
as unity~\cite{MO01a}.

\begin{figure}
\centering
\resizebox{0.75\columnwidth}{!}{\includegraphics{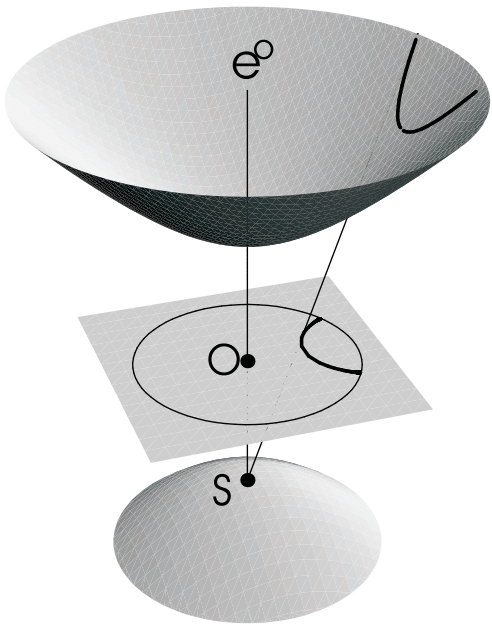}}
\caption{Outline of the unit hyperboloid and a geodesic on it. We
also show how a hyperbolic line is obtained in the unit disk via
stereographic projection taking the south pole as projection
center.}
\end{figure}

A simple calculation shows that if one uses
stereographic projection taking the south pole
$(-1, 0, 0)$ as projection center (see Fig.~2),
the projection of the point $(e^0, e^1, e^2)$
becomes in the complex plane
\begin{equation}
\label{defz}
z = \frac{e^1+ i e^2}{1 + e^0}=
\frac {E^{(-)}}{E^{(+)}} .
\end{equation}
The upper sheet of the unit hyperboloid is
projected into the unit disk, the lower sheet
into the external region, while the infinity
goes to the boundary of the unit disk.

The geodesics in the hyperboloid are intersections with the
hyperboloid of planes passing through the origin. Consequently,
hyperbolic lines are obtained from these by stereographic
projection and they correspond to circle arcs that orthogonally
cut the boundary of the unit disk.

It seems natural to consider the complex
variables in Eq.~(\ref{defz}) for both ambient
and substrate. In consequence, Eq.~(\ref{M1})
defines a transformation on the complex plane
${\mathbb{C}}$, mapping the point $z_s$ into
the point $z_a$ according to
\begin{equation}
\label{accion}
z_a = \Phi [\mathsf{M}_{as} , z_s] =
\frac{\beta^\ast +\alpha^\ast z_s}
{\alpha + \beta z_s} ,
\end{equation}
which is a bilinear (or M\"{o}bius) transformation.
The action of the multilayer can be seen as a
function $z_a = f(z_s)$ that can be appropriately
called the multilayer transfer function~\cite{MO02}.
The action of the inverse matrix
$\mathsf{M}_{as}^{-1}$ is $z_s = \Phi [
\mathsf{M}_{as}^{-1}, z _a]$.
One can show that the unit disk, the external
region and the boundary remain invariant under
the multilayer action.

For later purposes, we need the concept of distance
in the unit disk. To this end, it is customary to
define the cross ratio of four distinct points
$z_A$, $z_B$, $z_C$, and $z_D$ as the number
\begin{equation}
(z_A, z_B|z_C, z_D ) =
\frac{(z_A -z_C )/(z_B -z_C)}
{(z_A -z_D )/(z_B -z_D)} ,
\end{equation}
which is  real only when the four points lie
on a circle or a straight line. In fact,
bilinear transformations preserve this cross
ratio~\cite{PE70}.

Let now $z$ and $z^\prime$ be two points
that are joined by the hyperbolic line whose
endpoints on the unit circle are $E$ and $E^\prime$.
The hyperbolic distance between $z$ and $z^\prime$
is defined as
\begin{equation}
\label{hdis}
\mathrm{d}_\mathrm{H}(z, z^\prime )
= \frac{1}{2} | \ln (E, E^\prime|z, z^\prime ) | .
\end{equation}
This can be seen as arising from the usual Minkowski
distance in the unit hyperboloid (obtained through
geodesics) by stereographic projection~\cite{MI88}.
The essential point for our purposes here is
that bilinear transformations are isometries; i.e.,
they preserve this distance.

\section{Trace criterion for the classification of
multilayers}
\label{Trace}

Bilinear transformations constitute an
important tool in many branches of physics.
For example, in polarization optics they
have been employed for a simple classification
of polarizing devices by means of the concept
of eigenpolarizations of the transfer
function~\cite{AZ87}. The equivalent concept
in multilayer optics can be stated as the
field configurations such that $z_a = z_s
\equiv z_f$ in Eq.~(\ref{accion}), that is
\begin{equation}
z_f = \Phi [\mathsf{M}_{as} , z_f] ,
\end{equation}
whose solutions are
\begin{equation}
z_f = \frac{1}{2 \beta}
\left \{  -2 i \ \mathrm{Im}(\alpha) \pm
\sqrt{[\mathrm{Tr} ( \mathsf{M}_{as} )]^2 -4}
\right \} .
\end{equation}
These values $z_f$ are known as fixed
points of the transformation $\Phi$. The
trace of $\mathsf{M}_{as}$ provides then a
suitable tool for the classification of
multilayers~\cite{SA01}.

When $ [\mathrm{Tr} ( \mathsf{M}_{as} )] ^2
< 4$ the multilayer action is elliptic and
it has only one fixed point inside the unit
disk, while the other lies outside. Since
in the Euclidean geometry a rotation is
characterized for having only one invariant
point, this multilayer action can be
appropriately called a hyperbolic rotation.

When $ [ \mathrm{Tr} ( \mathsf{M}_{as} )]^2 > 4$
the multilayer action is  hyperbolic and it
has two fixed points both on the boundary of
the unit disk. The hyperbolic line joining these
two fixed points remains invariant and thus, by
analogy with the Euclidean case, this action will
be called a hyperbolic translation.

Finally, when $ [ \mathrm{Tr} (\mathsf{M}_{as}) ]^2 = 4$
the multilayer action is  parabolic and it
has only one (double) fixed point on the boundary
of the unit disk. This action will be called a
parallel displacement.

To proceed further let us note that by taking the
conjugate of $\mathsf{M}_{as}$ with any matrix
$\mathsf{C}\in $ SU(1,~1); i.e.,
\begin{equation}
\label{conjC}
\widehat{\mathsf{M}}_{as} = \mathsf{C} \
\mathsf{M}_{as} \ {\mathsf{C}}^{-1} ,
\end{equation}
we obtain another matrix of the same type, since
$\mathrm{Tr} (\widehat{\mathsf{M}}_{as}) =
\mathrm{Tr} (\mathsf{M}_{as})$. Conversely,
if two multilayer matrices have the same
trace, one can always find a matrix
$\mathsf{C}$ satisfying Eq.~(\ref{conjC}).

The fixed points of $\widehat{\mathsf{M}}_{as}$
are then the image by $\mathsf{C}$ of the
fixed points of $\mathsf{M}_{as}$. In
consequence, given any multilayer matrix
$\mathsf{M}_{as}$ one can always reduce it to
a $\widehat{\mathsf{M}}_{as}$ with one of the
following canonical forms:
\begin{eqnarray}
\label{Iwasa1}
\widehat{\mathsf{K}}(\varphi) & = &
\left [
\begin{array}{cc}
\exp (i\varphi/2) & 0 \\
0 & \exp (-i\varphi/2)
\end{array}
\right ]\ ,
\nonumber \\
\widehat{\mathsf{A}}(\chi) & = &
\left [
\begin{array}{cc}
\cosh (\chi/2) & i\, \sinh(\chi/2) \\
-i\, \sinh(\chi/2) & \cosh (\chi/2)
\end{array}
\right ]\ , \\
\widehat{\mathsf{N}}(\eta) & = &
 \left [
\begin{array}{cc}
1 - i\, \eta/2& \eta/2 \\
\eta/2 & 1+ i\, \eta/2
\end{array}
\right ]\ ,
\nonumber
\end{eqnarray}
that have as fixed points the origin
(elliptic), $+i$ and $-i$ (hyperbolic)
and $+i$ (parabolic), and whose physical
significance has been studied before~\cite{MO01b}.
The explicit construction of the family of
matrices $\mathsf{C}$ is easy: it suffices
to impose that $\mathsf{C}$ transforms the
fixed points of $\mathsf{M}_{as}$ into the
ones of $\widehat{\mathsf{K}}(\varphi)$,
$\widehat{\mathsf{A}}(\chi)$, or
$\widehat{\mathsf{N}}(\eta)$.

\begin{figure}
\centering
\resizebox{0.75\columnwidth}{!}{\includegraphics{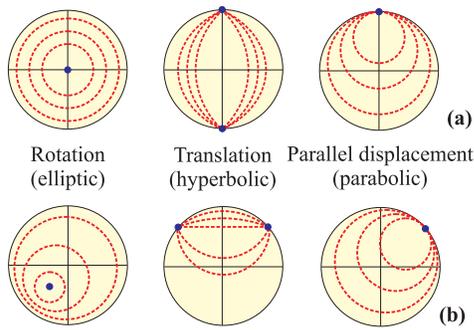}}
\caption{Plot of typical orbits in the unit disk for: (a)
canonical transfer matrices as given in Eq.~(\ref{Iwasa1}) and (b)
arbitrary transfer matrices.}
\end{figure}

The concept of orbit is especially appropriate for obtaining an
intuitive picture of these actions. We recall that given a point
$z$, its orbit is the set of points $z^\prime$ obtained from $z$
by the action of all the elements of the group. In Fig.~3.a we
have plotted typical orbits for each one of the canonical forms
$\widehat{\mathsf{K}}(\varphi)$, $\widehat{\mathsf{A}} (\chi)$,
and $\widehat{\mathsf{N}} (\eta)$.
For matrices $\widehat{\mathsf{K}}(\varphi)$ the orbits are circumferences
centered at the origin and there are no invariant hyperbolic
lines. For $\widehat{\mathsf{A}} (\chi)$, they are arcs of circumference
going from the point $ +i$ to the point $-i$ through $z$ and they
are known as hypercicles. Every hypercicle is equidistant [in the
sense of the distance (\ref{hdis})] from the imaginary axis, which
remains invariant (in the Euclidean plane the locus of a point at
a constant distance from a fixed line is a pair of parallel
lines). Finally, for $\widehat{\mathsf{N}} (\eta)$ the orbits are
circumferences passing through the point $+ i$ and joining the
points $z$ and $-z^\ast$ and they are known as horocycles: they
can be viewed as the locus of a point that is derived from the
point $+ i$ by a continuous parallel displacement~\cite{CO68}.

For a general matrix $\mathsf{M}_{as}$ the
corresponding orbits can be obtained by
transforming with the appropriate matrix
$\mathsf{C}$ the orbits described before.
In Fig.~3.b we have plotted typical examples
of such orbits for elliptic, hyperbolic,
and parabolic actions. We stress that once
the fixed points of the multilayer matrix
are known, one can ensure that $z_a$ will
lie in the orbit associated to $z_s$.

In the Euclidean plane any isometry is
either a rotation, a translation, or a
reflection. In any case, reflections are
the ultimate building blocks, since any
isometry can be expressed as the
composition of reflections. In this Euclidean
plane two distinct lines are either intersecting
or parallel. Accordingly, the composition of
two reflections in two intersecting lines
forming an angle $\varphi$ is a rotation of
angle $2 \varphi$ while the composition of two
reflections in two parallel lines separated
a distance $d$ is a translation of value $2 d$.

However, in the hyperbolic geometry induced in
the unit disk, any two distinct lines are either
intersecting (they cross in a point inside the
unit disk), parallel (they meet at infinity; i.e.,
at a point on the boundary of the unit disk),
or ultraparallel (they have no common points).
A natural question arises: what is the composition
of reflections in these three different kind of
lines? To some extent, the answer could be
expected: the composition is a rotation,
a parallel displacement, or a translation,
respectively. However, to gain further insights
one needs to know how to deal with reflections
in the unit disk. This is precisely the goal of
next Section.

\section{Reflections in the unit disk}

In the Euclidean plane given any straight
line and a point $P$ which does not lie on
the line, its reflected image $P^\prime$
is such that the line is equidistant from
$P$ and $P^\prime$. In other words, a
reflection is a special kind of isometry
in which the invariant points consist of all
the points on the line.

The concept of hyperbolic reflection is
completely analogous: given the hyperbolic
line $\Gamma$ and a point $P$, to obtain
its reflected image $P^\prime$ in $\Gamma$
we must drop a hyperbolic line $\Gammaç$
from P perpendicular to $\Gamma$ (such
a hyperbolic line exists and it is unique)
and extending an equal hyperbolic distance
[according to (\ref{hdis})] on the opposite
side of $\Gamma$ from $P$. In the unit
disk, this corresponds precisely to an
inversion.

To maintain this paper as self-containd
as possible, let us first recall some
facts about the concept of inversion.
Let $C$ be a circle with center $w$ and
radius $R$. An inversion on the circle $C$
maps the point $z$ into the point $z^\prime$
along the same radius in such a way that
the product of distances from the center
$w$ satisfies
\begin{equation}
| z^\prime - w | \
| z - w | = R^2 ,
\end{equation}
and hence one immediately gets
\begin{equation}
z^\prime = w + \frac{R^2}{z^\ast - w^\ast}
= \frac{R^2 + w z^\ast - w^\ast w}
{z^\ast - w^\ast} .
\end{equation}
If the circle $C$ is a hyperbolic line, it
is orthogonal to the boundary of the unit
disk and fulfills $w w^\ast = R^2 + 1$.
In consequence
\begin{equation}
\label{verinv}
z^\prime = \frac{w z^\ast -1}
{z^\ast - w^\ast} .
\end{equation}
One can check~\cite{PE70} that inversion
maps circles and lines into circles and lines,
and transforms angles into equal angles (although
reversing the orientation). If a circle $C^\prime$
passes through the points $P$ and $P^\prime$,
inverse of $P$ in the circle $C$, then $C$ and
$C^\prime$ are perpendicular.  Moreover, the
hyperbolic distance (\ref{hdis}) is invariant
under inversions. This confirms that inversions
are indeed reflections and so they appear as
the most basic isometries of the unit disk.

It will probe useful to introduce the conjugate
bilinear transformation associated with a
matrix $\mathsf{M}_{as}$ as [compare
with Eq.~(\ref{accion})]
\begin{equation}
\label{accionC}
z_a = \Phi^\ast [\mathsf{M}_{as}, z_s] =
\frac{\beta^\ast +\alpha^\ast z_s^\ast}
{\alpha + \beta z_s^\ast} .
\end{equation}
With this notation we can recast Eq.~(\ref{verinv}) as
\begin{equation}
\label{accionI}
z^\prime = \Phi^\ast [\mathsf{I}_w, z] ,
\end{equation}
where the matrix $\mathsf{I}_w \in$ SU(1,~1)
associated to the inversion is
\begin{equation}
\mathsf{I}_w =
\left [
\begin{array}{cc}
- i \ w^\ast/R & i/R \\
- i/R & i \ w/R
\end{array}
\right ] .
\end{equation}
The composition law for inversions can be
stated as follows: if $z^\prime =
\Phi^\ast [\mathsf{I}_w, z]$ and
$z^{\prime \prime} =
\Phi^\ast [\mathsf{I}_{w^\prime}, z^\prime]$
then
\begin{equation}
z^{\prime \prime} =
\Phi [\mathsf{I}_{w^\prime} \mathsf{I}_w^\ast , z] .
\end{equation}

To shed light on the physical meaning of the
inversion, assume that incoming and outgoing
fields are interchanged in the basic configuration
shown in Fig.~1. In our case, this is
tantamount to reversing the time arrow. It
is well know that given a forward-traveling
field $E^{(+)}$, the conjugate field $[E^{(+)}]^\ast$
represents a backward phase-conjugate wave of
the original field~\cite{ZE85}. In other words,
the time-reversal operation can be viewed
in this context as the transformation
\begin{equation}
z \mapsto \frac{1}{z^\ast} ,
\end{equation}
for both ambient and substrate variables; that
is, it can be represented by an inversion
in the unit circle. The transformed points
lie outside the unit circle because,
according to Eq.~(\ref{equivalencia}) this
time reversal transforms the upper sheet
into the lower sheet of the hyperboloid.

Moreover, by direct inspection is easy
to convince oneself that the matrix
relating these time-reversed fields is
precisely $\mathsf{M}_{as}^\ast$ and so
the action can be put as
\begin{equation}
(1/z_a)^\ast = \frac{\beta^\ast +\alpha^\ast
(1/z_s)^\ast}{\alpha + \beta (1/z_s)^\ast} ,
\end{equation}
which expresses a general property of the
time-reversal invariance in our model.

\section{Multilayer action as composition
of reflections}

As we have anticipated at the end
of Section~\ref{Trace} the composition
of two reflections gives a rotation,
a parallel displacement, or a translation,
accordingly the two hyperbolic lines
are intersecting, parallel, or ultraparallel,
respectively.

\begin{figure}
\centering
\resizebox{0.75\columnwidth}{!}{\includegraphics{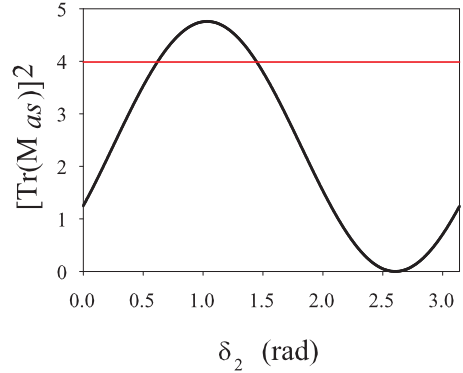}}
\caption{Plot of typical orbits in the unit disk for: (a)
canonical transfer matrices as given in Eq.~(\ref{Iwasa1}) and (b)
arbitrary transfer matrices.}
\end{figure}

To illustrate how this geometrical scenario
translates into practice, we consider an
optical system versatile enough so as it
could operate in the three regimes. To
this end, we choose a Fabry--Perot-like
system formed by two identical plates (each one
of them with fixed phase thickness $\delta_1$)
separated by a spacer of air with phase
thickness $\delta_2$. This is a symmetric
system for which $R_{as}$ and $T_{as}$
can be easily computed. By varying the spacer
phase-thickness $\delta_2$ we obtain the
values of $[\mathrm{Tr} (\mathsf{M}_{as})]^2$
shown in Fig.~4. In all the examples we take
as initial condition that in the substrate
$z_s = 0.4 \exp(- i \pi/3)$.

First, we take $\delta_2 = 3$ rad, so we are
in the elliptic case ($[\mathrm{Tr} (\mathsf{M}_{as})]^2
< 4$). From the computed values of $R_{as}$ and
$T_{as}$ one easily obtains the value $z_a =
-0.4447 + 0.4882 i$. The fixed point turns out
to be $z_f = - 0. 3114$ and, in consequence,
the multilayer action is a hyperbolic rotation
around the center $z_f$ of angle $2 \varphi$, as
indicated in Fig.~5. This multilayer action can
be viewed as the composition of reflections
in two hyperbolic lines $\Gamma_1$ and $\Gamma_2$
intersecting at $z_f$ and forming an angle $\varphi$.
The first inversion maps $z_s$ into the intermediate
point $z_{\mathrm{int}}$, which is mapped into
$z_a$ by the second inversion. Note that there
are infinity pairs of lines satisfying
these conditions, but chosen arbitrarily one
of them, the other is uniquely determined.
Moreover, once these lines are known, they
determine automatically the associated inversions.
\begin{figure}
\centering
\resizebox{0.75\columnwidth}{!}{\includegraphics{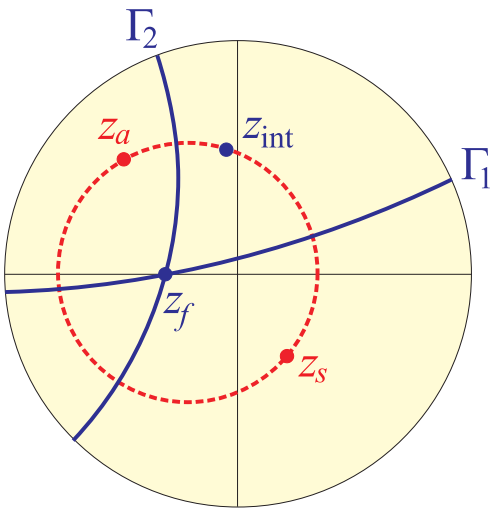}}
\caption{Decomposition of the multilayer action in terms of two
reflections in two intersecting lines for the same multilayer as
in Fig.~4 with $\delta_2 = 3$ rad (elliptic case).}
\end{figure}
\begin{figure}
\centering
\resizebox{0.75\columnwidth}{!}{\includegraphics{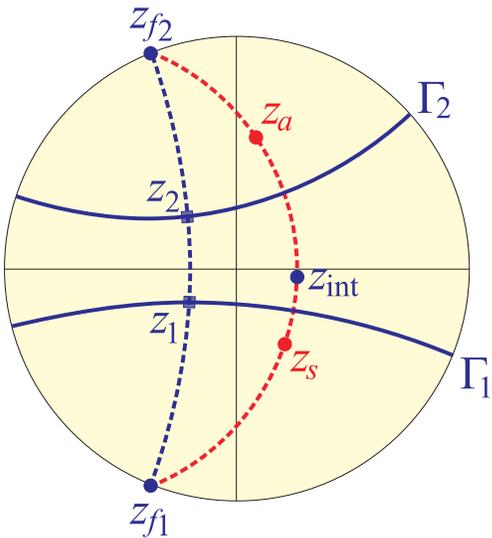}}
\caption{Decomposition of the multilayer action in terms of two
reflections in two ultraparallel lines for the same multilayer as
in Fig.~4 with $\delta_2 =1$ rad (hyperbolic case).}
\end{figure}
\begin{figure}
\centering
\resizebox{0.75\columnwidth}{!}{\includegraphics{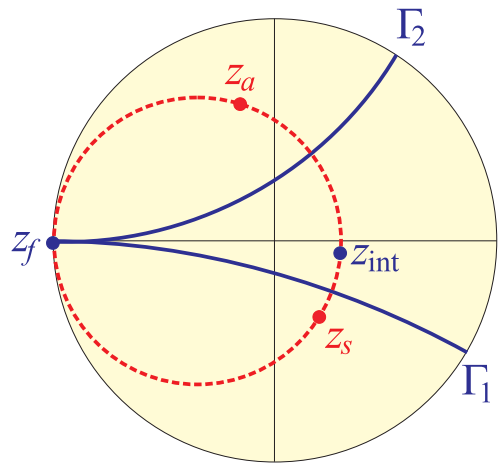}}
\caption{Decomposition of the multilayer action in terms of two
reflections in two parallel lines for the same multilayer as in
Fig.~4 with $\delta_2 = 0.4328$ rad (parabolic case).}
\end{figure}

Next, we take the spacer phase thickness $\delta_2 = 1$ rad, which
corresponds to the hyperbolic case ($[\mathrm{Tr}
(\mathsf{M}_{as})]^2 > 4$), and we get $z_a = 0.1567 + 0.4464 i$.
The fixed points are $z_{f1} = - 0.3695 - 0.9292 i$ and $z_{f2} =
- 0.3695 + 0.9292 i$. There are no invariant points in the unit
disk, but the hyperbolic line joining $z_{f1}$ and $z_{f2}$ is the
axis of the hyperbolic translation. In Fig.~6, we have plotted the
hypercicle passing through $z_a$ and $z_s$. The multilayer action
can be now interpreted as the composition of two reflections in
two ultraparallel hyperbolic lines $\Gamma_1$ and $\Gamma_2$
orthogonal to the translation axis. If  $\Gamma_1$ and $\Gamma_2$
intersect the hypercicle at the points $z_1$ and $z_2$, they must
fulfill
\begin{equation}
\label{tras}
d_{\mathrm{H}} (z_a, z_s) = 2
d_{\mathrm{H}} (z_1, z_2),
\end{equation}
in complete analogy with what happens in the
Euclidean plane. Once again, there are infinity
pairs of lines fulfilling this condition.

Finally, we take $\delta_2 = 0.4328$ rad, so we are in the
parabolic case ($[\mathrm{Tr} (\mathsf{M}_{as})]^2 = 4$), and
$z_a = - 0.1615 + 0.6220 i$. The (double) fixed point is $z_f = -
1$. In Fig.7 we have plotted the horocyle connecting $z_s$ and
$z_a$ and the fixed point. Now, we have the composition of two
reflections in two parallel lines $\Gamma_1$ and $\Gamma_2$ that
intersect at the fixed point $z_f$ and with the same constraint
(\ref{tras}) as before.

\section{Concluding remarks}

In this paper, we have proved a geometric
scenario to deal with multilayer optics. More
specifically, we have reduced the action
any lossless multilayer (no matter how complicated
it might be) to a rotation, a parallel displacement
or a translation, according to the magnitude
of its trace. These are the basic isometries
of the unit disk and we have expressed them as
the composition of two reflections in intersecting,
ultraparallel, or parallel lines. There is no
subsequent factorization in simpler terms so,
reflections are the most basic motions one
can find in the unit disk.

We hope that this approach will complement the more standard
algebraic method in terms of transfer matrices, and together they
will aid to obtain a better physical and geometrical feeling for
the properties of multilayers.

Finally, we stress that the benefit of this
formulation lies not in any inherent advantage
in terms of efficiency in solving problems in
layered structures. Rather, we expect that the
formalism presented here could provide a general
and unifying tool to analyze multilayer performance
in a way closely related to other fields of physics,
which seems to be more than a curiosity.

\section*{Acknowledgments}
We thank J. Zoido for his help in computing
some of the figures of this paper.

Corresponding author Luis L. S\'anchez-Soto
e-mail address is lsanchez@fis.ucm.es.


\begin{thebibliography}{99}

\bibitem{SC97}
B. F. Schutz,
\textit{Geometrical methods of Mathematical Physics}
(Cambridge University Press, Cambridge, 1997).

\bibitem{MO99a}
J. J. Monz\'{o}n and L. L. S\'{a}nchez-Soto,
``Lossless multilayers and Lorentz transformations:
more than an analogy,''
Opt. Commun. \textbf{162}, 1-6 (1999).

\bibitem{MO99b}
J. J. Monz\'{o}n and L. L. S\'{a}nchez-Soto,
``Fully relativisticlike formulation of multilayer optics,''
J. Opt. Soc. Am. A \textbf{16}, 2013-2018 (1999).

\bibitem{CO68}
H. S. M. Coxeter,
\textit{Introduction to Geometry}
(Wiley, New York, 1969).

\bibitem{YO02}
T. Yonte, J. J. Monz\'{o}n, L. L. S\'{a}nchez-Soto,
J. F. Cari\~{n}ena, and C. L\'opez-Lacasta,
``Understanding multilayers from a geometrical viewpoint,''
J. Opt. Soc. Am. A \textbf{19}, 603-609 (2002).

\bibitem{MO02}
J. J. Monz\'{o}n, T. Yonte, L. L. S\'{a}nchez-Soto, and J. F.
Cari\~{n}ena, ``Geometrical setting for the classification of
multilayers,'' J. Opt. Soc. Am. A \textbf{19}, 985-991 (2002).

\bibitem{AZ87}
R. M. A. Azzam and N. M. Bashara,
\textit{Ellipsometry and Polarized Light}
(North-Holland, Amsterdam, 1987).

\bibitem{HA96}
D. Han, Y. S. Kim, and M. E. Noz,
``Polarization optics and bilinear representations
of the Lorentz group,"
Phys. Lett. A \textbf{219}, 26-32 (1996).

\bibitem{KO65}
H. Kogelnik,
``Imaging of optical modes --resonators
with internal lenses,''
Bell Syst. Techn. J. \textbf{44}, 455-494 (1965).

\bibitem{NA98}
M. Nakazawa, J. H. Kubota, A. Sahara, and
K. Tamura,
``Time-domain ABCD matrix formalism for
laser mode-locking and optical pulse transmission,''
IEEE J. Quant. Electron. \textbf{QE34}, 1075-1081 (1998).

\bibitem{ME91}
R. Melter, A. Rosenfeld, and P. Bhattacharya,
\textit{Vision Geometry}
(American Math. Soc., Providence, 1991).

\bibitem{DU81}
K. A. Dunn, ``Poincar\'e group as reflections in
straight lines," Am. J. Phys. \textbf{49}, 52-55 (1981).

\bibitem{Snell}
When ambient ($0$) and substrate ($m+1$) media are
different, the angles $\theta _{0}$ and
$\theta _{m+1}$ are conected by Snell
law $n_0\sin \theta_0= n_{m+1}\sin
\theta_{m+1}$, where $n_{j}$ denotes the
refractive index of the $j$th medium.

\bibitem{MO99c}
J. J. Monz\'{o}n and L. L. S\'{a}nchez-Soto,
``Origin of the Thomas rotation that arises
in lossless multilayers,''
J. Opt. Soc. Am. A  \textbf{16}, 2786-2792 (1999).

\bibitem{MO01a}
J. J. Monz\'{o}n and  L. L. S\'{a}nchez-Soto,
``A simple optical demonstration of geometric phases
from multilayer stacks: the Wigner angle as an
anholonomy,''
J. Mod. Opt. \textbf{48}, 21-34 (2001).

\bibitem{PE70}
D. Pedoe,
\textit{A course of Geometry}
(Cambridge Universtiy Press, Cambridge, 1970).

\bibitem{MI88}
A. Mischenko and A. Fomenko,
\textit{A Course of Differential Geometry
and Topology} (MIR, Moscow, 1988), Sec. 1.4.

\bibitem{SA01}
L. L. S\'anchez-Soto, J. J. Monz\'on,
T. Yonte, and J. F. Cari\~{n}ena,
``Simple trace criterion for classification of multilayers,"
Opt. Lett. \textbf{26}, 1400-1402 (2001).

\bibitem{MO01b}
J. J. Monz\'{o}n, T. Yonte, and L. L. S\'{a}nchez-Soto,
``Basic factorization for multilayers,''
Opt. Lett. \textbf{26}, 370-372 (2001).

\bibitem{ZE85}
B. Ya. Zel'dovich, N. F. Pilipetsky, and V. V. Shkunov.
\textit{Principles of Phase Conjugation}
(Springer-Verlag, Berlin, 1985).

\end{thebibliography}
\end{document}